\documentclass[reprint,twocolumn,prl,amsmath,amssymb,longbibliography,aps,superscriptaddress]{revtex4-1}

\pdfoutput=1

\usepackage{graphicx}
\usepackage{amsmath,latexsym,tabularx}
\usepackage{xcolor}
\usepackage[%
  colorlinks=true,
  urlcolor=blue,
  linkcolor=blue,
  citecolor=blue
]{hyperref}

\newcommand{\affilLL}[0]{Lincoln Laboratory, Massachusetts Institute of Technology, Lexington, Massachusetts 02421, USA}
\newcommand{\affilMIT}{Massachusetts Institute of Technology, Cambridge, Massachusetts 02139, USA}

\begin{document}

\title{Distance scaling of electric-field noise in a surface-electrode ion trap}

\author{J. A. Sedlacek}
\email[]{jonathon.sedlacek@ll.mit.edu}
\affiliation{\affilLL}

\author{A. Greene}
\affiliation{\affilLL}
\affiliation{\affilMIT}

\author{J. Stuart}
\affiliation{\affilLL}
\affiliation{\affilMIT}

\author{R. McConnell}
\affiliation{\affilLL}

\author{C. D. Bruzewicz}
\affiliation{\affilLL}

\author{J. M. Sage}
\email[]{jsage@ll.mit.edu}
\affiliation{\affilLL}

\author{J. Chiaverini}
\email[]{john.chiaverini@ll.mit.edu}
\affiliation{\affilLL}

\date{\today}

\begin{abstract}

We investigate anomalous ion-motional heating, a limitation to multi-qubit quantum-logic gate fidelity in trapped-ion systems, as a function of ion-electrode separation.  Using a multi-zone surface-electrode trap in which ions can be held at five discrete distances from the metal electrodes, we measure power-law dependencies of the electric-field noise experienced by the ion on the ion-electrode distance~$d$.  We find a scaling of approximately~$d^{-4}$ regardless of whether the electrodes are at room temperature or cryogenic temperature, despite the fact that the heating rates are approximately two orders of magnitude smaller in the latter case.  Through auxiliary measurements using application of noise to the electrodes, we rule out technical limitations to the measured heating rates and scalings.  We also measure frequency scaling of the inherent electric-field noise close to~$1/f$ at both temperatures.  These measurements eliminate from consideration anomalous-heating models which do not have a~$d^{-4}$ distance dependence, including several microscopic models of current interest.

\end{abstract}

\maketitle

Trapped atomic ions manipulated using electromagnetic fields are a promising system for large-scale quantum information processing (QIP)~\cite{Leibfried2003}, but the fidelity of multi-qubit operations based on the Coulomb interaction is limited by ion heating and motional-state decoherence caused by electric-field noise.  This noise is observed to be significantly larger in amplitude than the expected thermal (Johnson) noise of the voltage on the trap electrodes and associated circuitry~\cite{theBible,Turchette2000}, and has therefore been termed anomalous.  Anomalous motional heating has been suspected, and observed in a few experiments~\cite{Turchette2000,PhysRevLett.97.103007,hite_mckay_kotler_leibfried_wineland_pappas_2017,wunderlich_arXiv_2017}, to be strongly dependent on the spatial separation between the ions and the electrode surfaces.  The reduction of this separation, which results in higher trap frequencies, and thus stronger ion-ion interactions, will be beneficial for increasing the speed of trapped-ion quantum logic based either on optical~\cite{Inns:HiFi2qubit:NatPhys:08,PhysRevLett.117.060504,PhysRevLett.117.060505} or magnetic-field-gradient~\cite{PhysRevLett.117.140501} excitation.  Additionally, high-rate, high-fidelity ion-state readout across an array can be enabled by trapping ions close to detectors integrated into trap chips~\cite{KaranMehtaThesis,Slichter:17}.   It is therefore of paramount importance to carefully determine how electric-field noise varies with ion-electrode distance~$d$ in potentially scalable systems.

Theories put forth to explain anomalous heating suggest power-law scalings of the electric-field spectral density (proportional to the heating rate) of $d^{-\beta}$ with the heating-rate exponent~$\beta$ in the range of~$0$ to~$8$~\cite{Brownnutt2015}.  An early model suggested to explain anomalous heating depends on small ``patches'' of the electrode surface, each with different electric properties, such as local work function, that lead to contact potentials between the patches, and therefore spatial electric-field variation above the surface~\cite{theBible}.  Time variation of these potentials due to thermal noise or other excitation can lead to ion heating.  For patches small in lateral extent compared to $d$, a scaling exponent $\beta=4$ is expected~\cite{PhysRevA.80.031402,PhysRevA.84.053425}.

While anomalous heating has been observed in many experiments in many types and sizes of trapping structures~\cite{Brownnutt2015}, systematic and uncontrolled variations between experiments, materials, geometries, and techniques have precluded direct comparison to obtain reliable scalings without invoking assumed scalings in other quantities (cf.~\cite{Turchette2000} in which scaled traps were measured in separate experiments).  The few experiments that have been performed to explore this scaling in a single device or experiment have been performed either:  using trap geometries that are not easily scalable~\cite{PhysRevLett.97.103007,hite_mckay_kotler_leibfried_wineland_pappas_2017}, which may limit their technological relevance for QIP; or in the limit of high ion temperature~\cite{wunderlich_arXiv_2017} with potentially limited bearing on the few-motional-quanta level required for high-fidelity quantum logic.  Temperature dependence of the exponent, which may help elucidate underpinnings of electric-field noise, has not been previously measured in a trapped-ion context, to our knowledge.

Here we measure the dependence of anomalous heating on ion-electrode separation in a single trap with multiple zones of varying trapping distances.  We use a surface-electrode design~\cite{NIST:SET:QIC:05}, which has great potential for scaling to large arrayed structures while employing integrated control and readout technologies required for large-scale QIP~\cite{Mehta_NNano_2016,Lekitsch2017}.  We measure heating rates from the ground state of the ion's motion in the trap potential using the motional-sideband-ratio~\cite{PhysRevLett.75.4011} and Rabi-flopping~\cite{RoweQIC2002} techniques, robust methods for measuring heating in the low-excitation (few-phonon) limit that assume only a thermal motional state of the ion.  We measure the noise as a function of~$d$, and as a function of frequency at one value of~$d$, for two electrode temperatures, $295$~K and $5$~K.  We observe that the distance and frequency scalings are independent of temperature.  Additionally, we apply voltage noise to the electrodes to verify that the nominal heating-rate scaling is not due to technical noise (i.e. noise from pickup in the wires or generated by the sources and circuitry used to apply potentials to the electrodes).  We observe a different, expected distance dependence of heating in this circumstance, suggesting that the intrinsic noise appears to originate from the electrode surface itself.

\begin{figure}[t b !]

\includegraphics[width = \columnwidth]{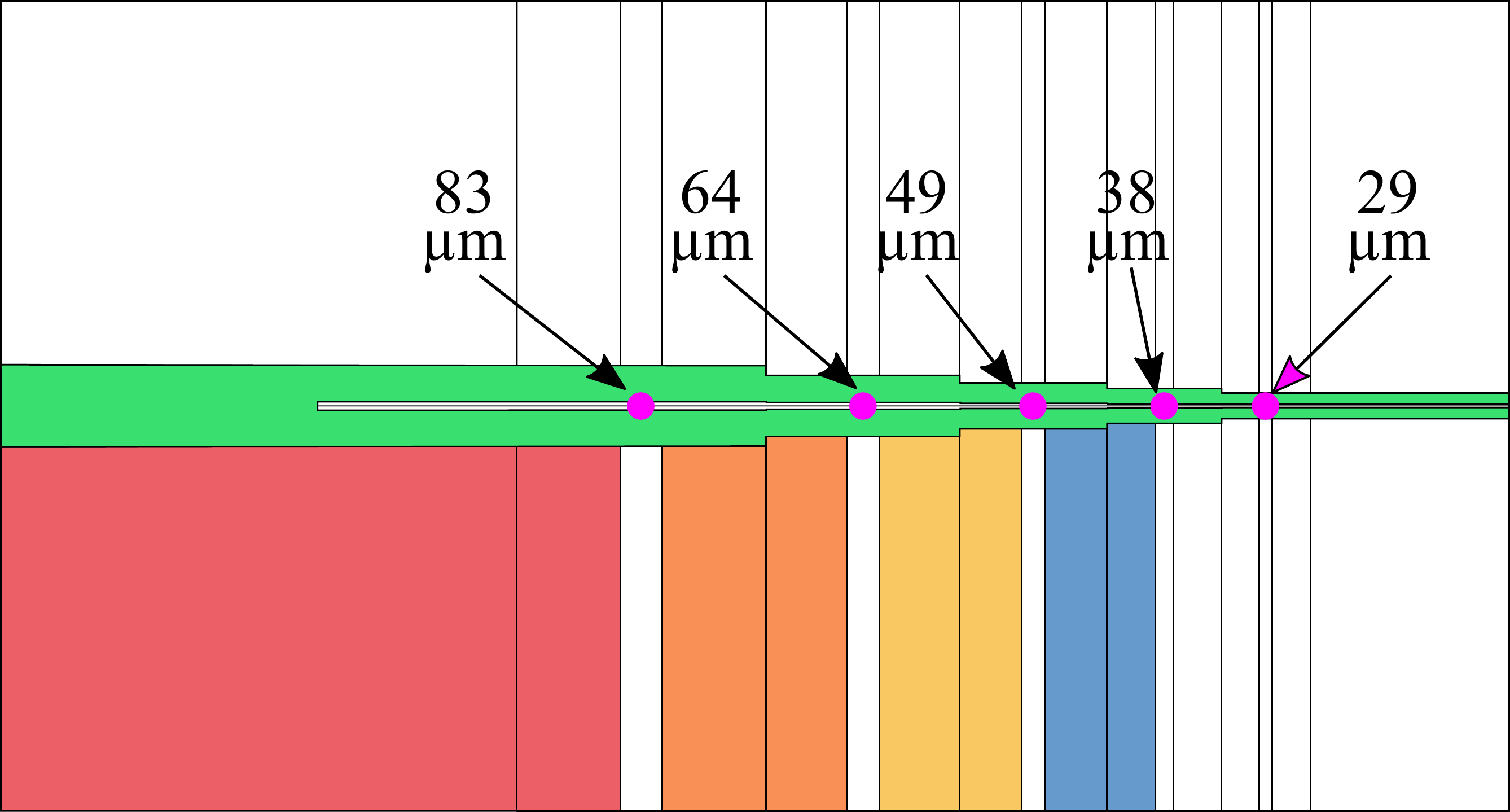}\\
\vspace{0.2 cm}
\includegraphics[width = \columnwidth]{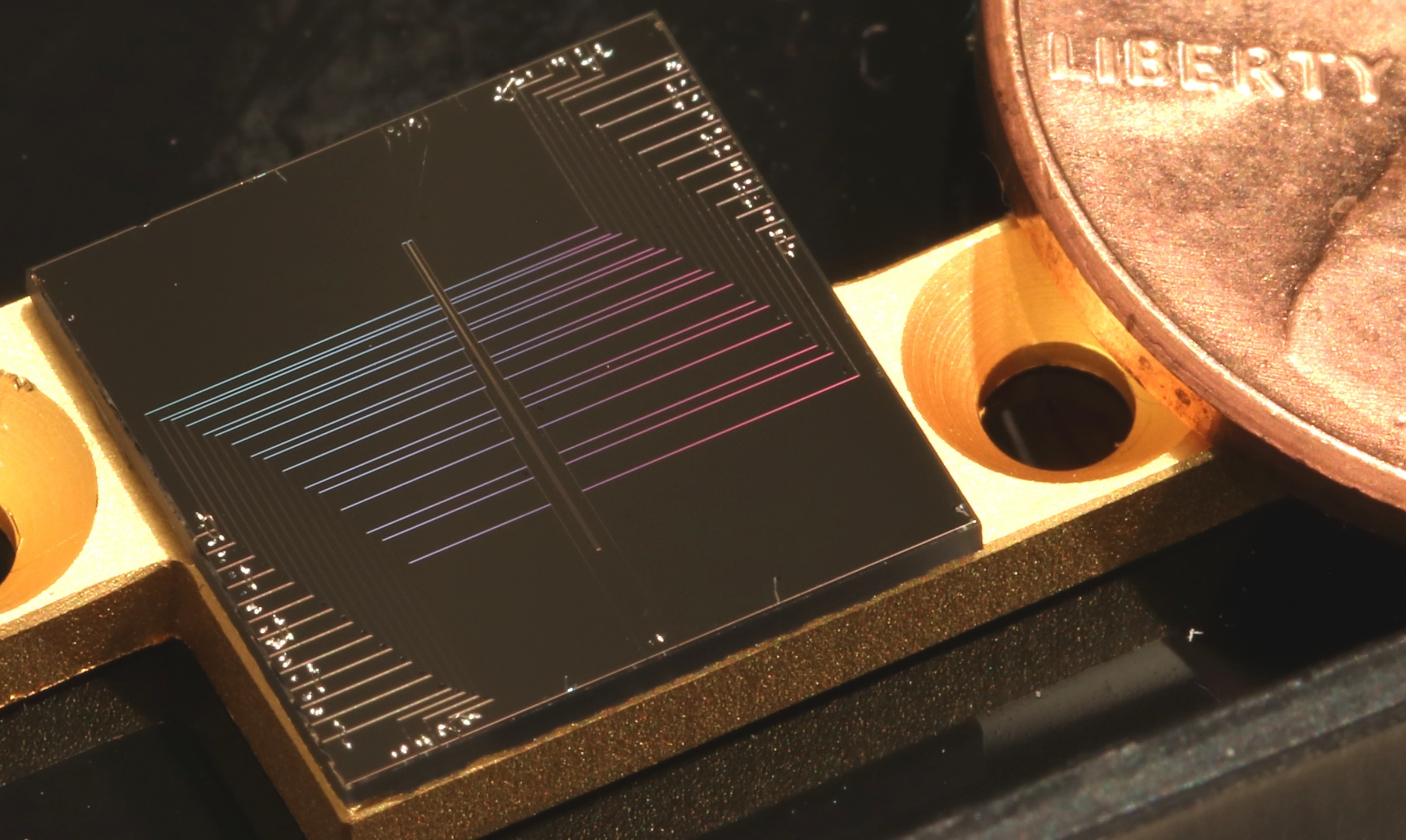}
\caption{Variable-height surface-electrode trap.  The upper figure shows the layout of the electrodes in the central region of the chip where the ions are trapped.  Purple dots represent the centers of the various zones axially separated along the common radio-frequency electrode (highlighted in green).  The ions are trapped above these locations at the distance labeled in each of the respective zones.  Sets of electrodes to which noise was added to determine the technical-noise scaling are denoted by color; electrodes in each set were connected together on-chip.  The lateral extent of this figure is approximately~$7.6$~mm.  The lower figure is a photograph of the 1~cm$^{2}$ chip, after demounting from the chamber and removal of wire-bonds, with a U.S. one-cent piece for scale.  The different zones can be seen in the center of the chip, largest to smallest from bottom to top in this image.}
\label{fig:electrodes}
\end{figure}

We employ a surface-electrode trap made from a $2$~$\mu$m-thick sputtered niobium electrode layer deposited on a sapphire substrate.  Electrodes were defined in the metal film via optical lithography and plasma etching.  The electrode design creates a linear Paul trap of varying ion-surface distance in five different zones along the trap axis (cf. Fig.~\ref{fig:electrodes}).  The ion-surface distance is determined by the lateral size and spacing of the radio-frequency (RF) electrodes in this structure~\cite{PhysRevA.78.033402,PhysRevA.78.063410}, and the five zones produce trapping minima at nominal ion-surface separations of $83$, $64$, $49$, $38$, and $29$~$\mu$m.  The uncertainty in~$d$ is approximately~1~$\mu$m, primarily due to trap potential simulation precision.  Ions can be trapped in all of these zones using a combination of RF fields and static potentials applied to subsets of the segmented control electrodes that axially define each zone.

For heating-rate measurements, a single $^{88}$Sr$^{+}$ ion is held in one of these five zones, and laser beams used to cool, manipulate, and read out the ion's electronic state are directed parallel to the trap surface, focused on the ion.  The trap is housed in a cryogenic vacuum system (described previously~\cite{Chiaverini2014}) in which ultra-high vacuum conditions can be sustained while the trap chip temperature is maintained at room temperature or $5$~K.  Loading is accomplished from a two-dimensional magneto-optical trap of neutral strontium atoms remotely located and accelerated to the trap where they are photoionized~\cite{bruzewicz2016scalable}.  Doppler cooling on a strong dipole-allowed transition and resolved-sideband cooling on the optically addressed $^{2}$S$_{1/2}$~$\rightarrow$~$^{2}$D$_{5/2}$ transition bring the ion to the ground state of the axial mode of vibration in the trap.  Sideband spectroscopy~\cite{PhysRevLett.75.4011} or carrier Rabi flopping~\cite{RoweQIC2002} on the optical transition is then used to measure the axial-mode ground state occupation after varying post-cooling delays.

Sideband spectroscopy, where the mode occupation is determined from the ratio of the red and blue motional-sideband amplitudes, is used for all measurements except the room-temperature distance scaling.  In the former case, the ground state occupation probability was typically $>0.95$ after sideband cooling.  In the latter case, the Rabi-flopping technique, where the mode occupation is determined by fitting electronic-carrier-transition Rabi flopping curves, was used due to the lower uncertainty of this method for higher motional excitation (a few to $\sim10$~phonons; here, for the highest heating rates, the initial mode occupation was $\bar{n}\leq 2$).  We crosschecked the two methods against each other and see general agreement, with the Rabi-flopping method producing systematically higher heating-rate values by 30\%, consistent with previous work~\cite{PhysRevA.89.062308}.  As the entire distance scaling data set at $295$~K was taken in this manner, the analysis of the scaling exponent of the heating rate is insensitive to this systematic uncertainty in measurement, giving us confidence in the exponent at much better than the 30\% level, ultimately limited by statistical uncertainty in the measurements.

For these experiments, we employ low-pass filtering below the trap frequency on all control electrodes.  Additionally, we band-pass filter the applied RF signal and, in every trap zone, we compensate stray fields so as to minimize the pseudopotential gradient along the axial direction and thus eliminate heating from noise near the RF frequency~\cite{arxiv_noise_paper,Blakestad2009}.

\begin{figure}[t b !]
\includegraphics[width = \columnwidth]{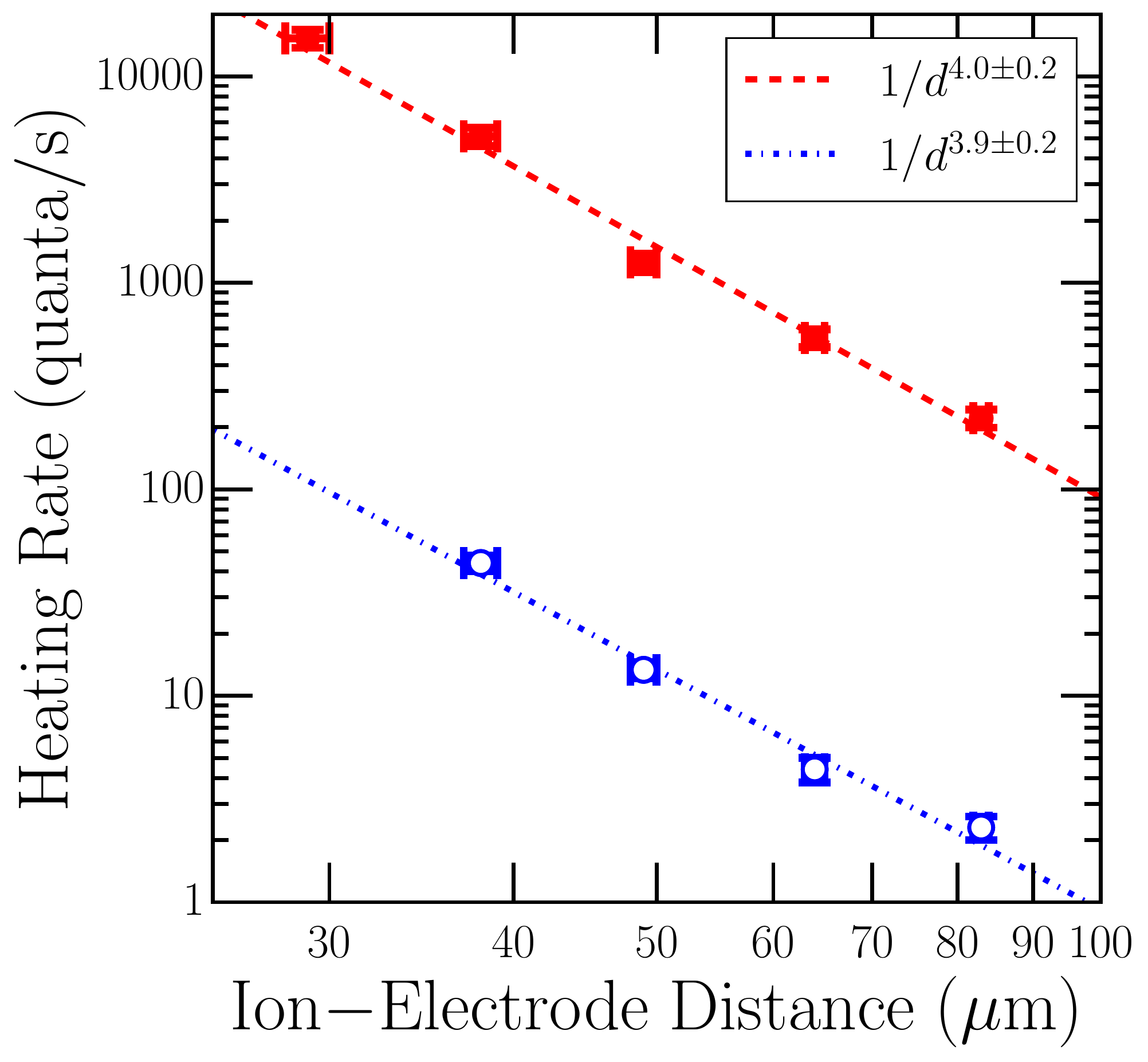}
\caption{Scaling of ion heating rate with ion-electrode distance at two different temperatures.  Heating rates were measured on a single-ion axial mode at $850$~kHz in all cases; the red, solid [blue, open] points are for a chip temperature of $295(1)$~K [$5(1)$~K].  The lines are fits to the data in each set, and the exponents and uncertainties denoted in the legend are determined from these fits.  Error bars are standard uncertainties in the heating rates as propagated through the sideband-ratio measurements.}
\label{fig:distance_scaling}
\end{figure}

To determine the scaling of electric-field noise with~$d$, we performed measurements of the heating rate at a fixed axial frequency of $850$~kHz in different trap zones.  (The behavior of the ion in the 29 um zone when the chip was at a temperature of $5$~K was atypical, leading us to believe uncompensated stray fields were often present in this circumstance; we therefore did not obtain heating-rate data for this case.)  These data are plotted in Fig.~\ref{fig:distance_scaling} for~$295$~K and~$5$~K chip temperatures.  While the overall heating rates are $\sim\nobreak100\times$ smaller at $5$~K than at $295$~K, the distance scalings are the same within error, with values of $\beta$ of~$3.9(2)$ and~$4.0(2)$, respectively, where the number in parentheses is the standard uncertainty in the last digit.  We note that the niobium electrode metal is superconducting at $5$~K, and the heating-rate scaling is therefore unchanged above and below the transition temperature $T_{C}$; the insensitivity to bulk superconductivity is not unexpected in light of previous data demonstrating no significant difference in heating-rate amplitude just above and below $T_{C}$~\cite{Wang2010,Chiaverini2014}.

\begin{figure}[t b !]
\includegraphics[width = 0.95\columnwidth]{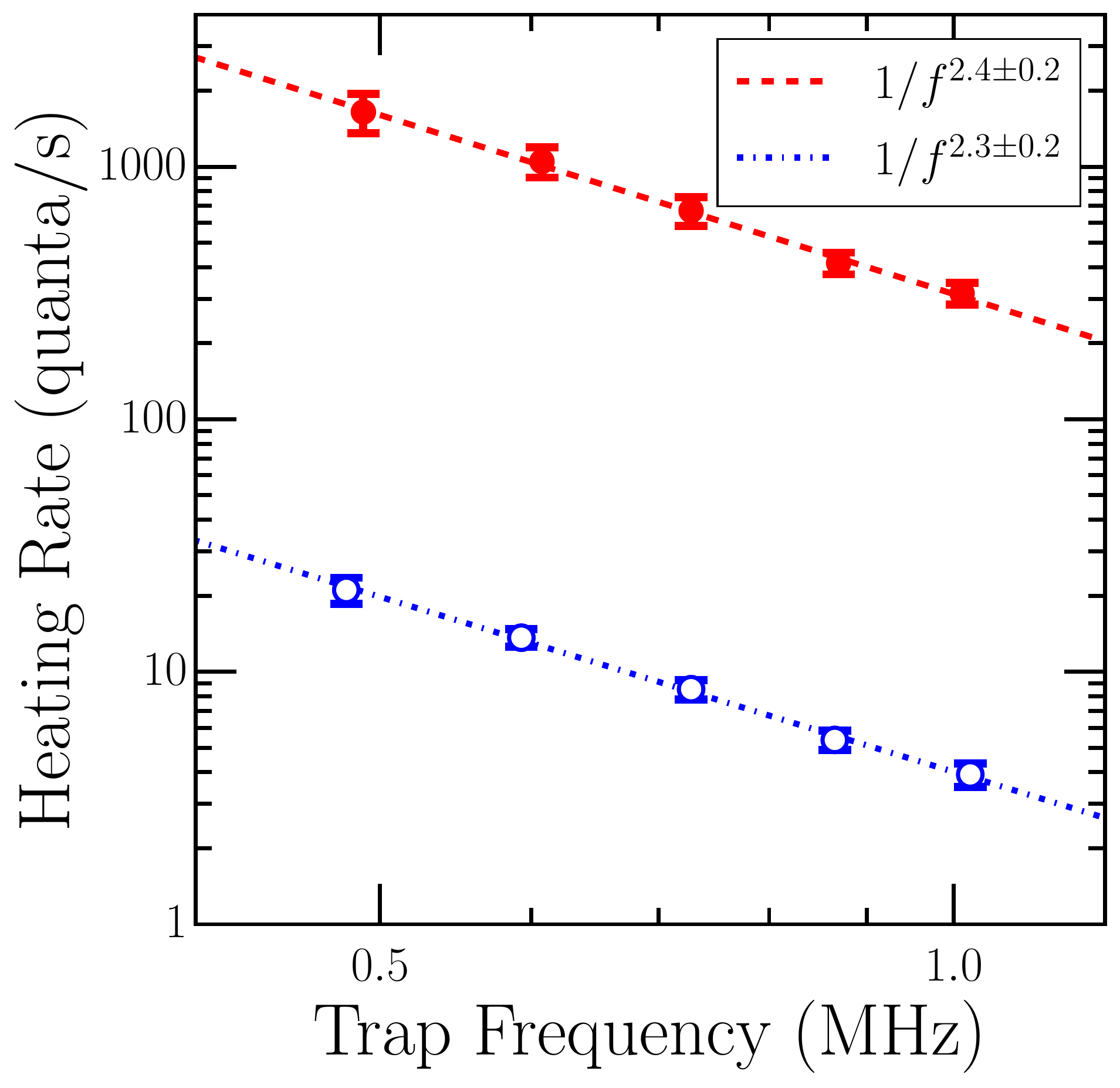}
\caption{Frequency scaling of the heating rate of the ion's axial mode as a function of trap frequency at two different temperatures with $d=64$~$\mu$m.  Red, solid [blue, open] points are taken at $295(1)$~K [$5(1)$~K]. The lines are power-law fits to the data points taken at each temperature, and the determined exponents and uncertainties are denoted in the legend.}
\label{fig:freqScaling}
\end{figure}

Measurement of the frequency dependence of the ion heating rate can also aid in determining the electric-field noise source due to the different dependencies predicted by particular models.  The frequency dependence of the ion heating rate was measured in one zone ($d=64$~$\mu$m), by varying the axial trap frequency, at both room and low temperature; these data are presented in Fig.~\ref{fig:freqScaling}.  The heating rate $\dot{\bar{n}}$ is proportional to the electric-field spectral density $S_{E}(\omega)$ as

\begin{equation}
\dot{\bar{n}} = \frac{e^{2}}{4 m \hbar \omega}S_{E}(\omega).
\label{equation_heating}
\end{equation}

\noindent Here $e$ and $m$ are the ion's charge and mass, respectively, $\hbar$~is the reduced Planck's constant, and~$\omega$ is the trap angular frequency ($=2\pi\times\nobreak f$, for frequency $f$).  Measurement of the heating rate thus gives the frequency scaling of $S_{E} (\omega)$.  We observe frequency scalings of the electric-field spectral density as $f^{-1.4(2)}$ and $f^{-1.3(2)}$ at room and low temperature, respectively.  This is consistent with $1/f$-like frequency dependence seen in several other experiments~\cite{Brownnutt2015}, and the fact that it is unchanged over this temperature range suggests that the mechanisms responsible for heating at high and low temperature are similar or related.  The observed distance scaling also supports this interpretation.

In the regime where technical noise or Johnson noise dominates, the heating rate should scale as $\sim 1/D_{i,j}^2$ \cite{Leibrandt2007}, where $D_{i,j}$ is the characteristic distance of electrode~$j$ for fields along direction~$i$ at the ion location.  In the multi-segment, variable-height trap we use here, characteristic distances for electrodes seen by the ion in each of the zones do not directly correspond to the ion-electrode distances $d$ in those zones, as the staggered shape of the RF electrodes, the lack of constant scaling of all dimensions of the electrodes among zones, and the presence of connected control electrodes of differing sizes in neighboring zones affect, in a non-symmetric manner, the field produced by each electrode.  Through boundary-element-method potential simulation of our trap, we have calculated these characteristic distances and their effect on distance scaling.  For technical or Johnson noise on the electrodes, including electromagnetic pickup, these characteristic distances lead to an expected approximate-power-law scaling exponent $\beta=2.4(3)$ when considering the ion-electrode distances used here.  The difference from perfect power-law-scaling, here quantified as an uncertainty in the exponent, is due to the unequal scalings of the electrode axial dimensions among zones.  Uncertainty in the ion's position ($\sim\nobreak1$~$\mu$m) leads to an expected heating-rate contribution uncertainty correlated to this position.

\begin{figure}[t b !]
\includegraphics[width = \columnwidth]{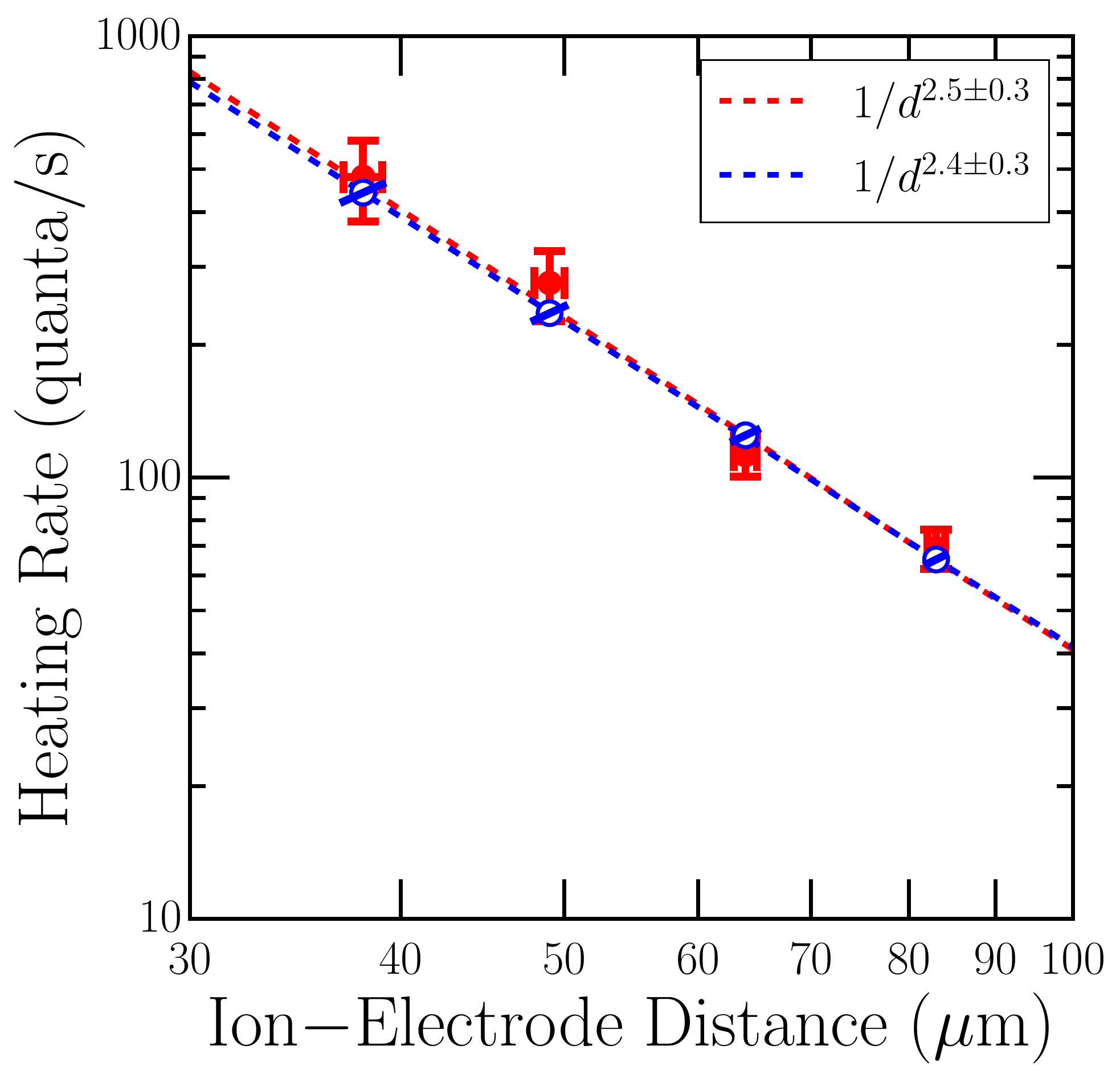}
\caption{Heating rate contributions due to the injection of technical noise. Noise is injected onto adjacent electrodes as described in the text.  The solid points shown in red are experimental data, all taken at $5(1)$~K.  A background heating rate (i.e. with no added noise) is measured at each distance and is subtracted from the the total heating rate.  The blue, open circles are the expectation of the increased heating rate due to calculation of the characteristic distances of the various electrode sets (see text).  The diagonal bars on the latter data represent the (correlated) uncertainty range of heating rates expected due to the uncertainty in ion position.  The legend depicts the fitted exponent scalings and uncertainties.}
\label{fig:NoiseScaling}
\end{figure}

To experimentally investigate the distance scaling of injected noise, we follow a procedure outlined briefly here; it will be described in detail in a future publication~\cite{arxiv_noise_paper}.  White noise, over a $10$~Hz to $2$~MHz range and at a level $>40$~dB higher than the intrinsic noise measured on the leads going to the control electrodes, is injected onto a pair of electrodes that are connected electrically on the chip (some electrodes are shared between zones to minimize required leads; cf. Fig.~\ref{fig:electrodes}).  The contribution to the heating rate from externally injected noise is determined for each zone.  In Fig.~\ref{fig:NoiseScaling}, we plot the results of this measurement along with the scaling determined from the calculated characteristic distances.  As can be seen, the measured heating-rate scaling of $\beta=2.5(3)$ is equivalent to the expected scaling within error, and is significantly different from the value $\beta=3.9(2)$ observed without noise injection.  This confirms that the scaling measured without injected noise is not limited by technical or Johnson noise, nor by pickup within the chamber.  We also note that the measured distance scaling without injected noise eliminates black-body radiation (BBR) as a limiting source of electric-field noise; close to an electrode surface, BBR will lead to a heating rate with scaling exponent~$\beta$ in the range of $2$--$3$~\cite{Brownnutt2015}.

The measured distance scaling puts limits on theories proposing microscopic origins of anomalous heating, assuming that the mechanism for heating is the same in each site.  A general patch potential model, agnostic to mechanism within the patches, predicts $\beta=4$ for patches much smaller than $d$, consistent with what is seen here~\cite{PhysRevA.80.031402,PhysRevA.84.053425,Brownnutt2015}.  Our data are thus consistent with patch size much smaller than $\sim\nobreak30$~$\mu$m, therefore eliminating electrode- or chip-sized patches, i.e. uniform fluctuations on a large scale, for which one would predict $\beta=2$.  We expect the crystalline grains in the polycrystalline sputtered niobium film in which the electrodes are defined to be not larger than $200$~nm in size ($\ll d$), but it is not possible to verify that these grains correspond to patches without making measurements at much smaller~$d$.  The adatom diffusion model~\cite{theBible,Brownnutt2015} predicts $\beta=6$ for noise measured parallel to the surface of planar electrodes, and thus would appear to be eliminated over the entire temperature range; this is true also for extensions that depend on diffusion between small patches that allow for spatio-temporal variation in induced dipole moment~\cite{Brownnutt2015,PhysRevA.95.033407}.

Models with adatom dipoles that fluctuate in size~\cite{PhysRevA.87.023421} predict $\beta=4$ as seen here, but the frequency scaling of the noise expected in this case is flat, corresponding to $f^{-1}$ scaling for the heating rate by Eq.~\ref{equation_heating}, inconsistent with the scaling seen here at both temperatures, and with almost every other measurement citing a frequency scaling~\cite{Brownnutt2015}.  Two-level fluctuators (of which adatom dipoles are potentially a subset) have also been suggested~\cite{RevModPhys.53.497,Brownnutt2015}, particularly in light of their potential relevance to electronic noise in conductors and $1/f$-type predictions.  This model also predicts $\beta=4$ for planar electrodes, and so is consistent with all the data presented here.  This model is however not supported by more extensive measurements of temperature and frequency scaling~\cite{Labaziewicz2008,Bruzewicz2015}.

One model that is consistent with the measurements performed here is based on lossy dielectric films atop the electrodes~\cite{kumph_NJP_2016}.  This model predicts electric-field noise proportional to~$1/f$ and to temperature as $T$, with $\beta=4$.  There is also dependence on loss tangent and dielectric constant of the thin film, properties whose frequency and temperature dependence could lead to the overall dependencies seen previously (e.g.~\cite{Bruzewicz2015}).  We also have evidence from X-ray photoelectron spectroscopy of fabricated traps that thin layers of dielectric materials, namely hydrocarbons and, in some materials, oxides, are present on the electrode surfaces.  Further investigation of these film material properties on fabricated trap electrodes, as well as measurements of heating rates as a function of film thickness and composition, is required to more carefully determine the relevance of this mechanism.

The results of a recent room-temperature distance dependence measurement in a gold trap~\cite{wunderlich_arXiv_2017} are similar to those observed here in a niobium trap, and the former experiment also measured the heating of an axial mode parallel to the trap surface.  Besides lending more support to the conclusions drawn above pertaining to model viability, this also suggests that material insensitivity seen previously in temperature dependence of electric-field noise~\cite{Chiaverini2014} is a more general phenomenon, one that any microscopic theory explaining anomalous heating must address.

These results suggest that scaling to smaller trap structures to perform high-fidelity quantum logic gates at higher rates will indeed be impractical without further mitigation of anomalous heating.  As a result, understanding the origins of anomalous ion heating remains of foremost importance.  Our measurement of the ion-electrode-distance scaling of electric-field noise strongly constrains proposed models for the microscopic mechanism(s) behind anomalous ion heating.  As this scaling appears to be unchanged between 295~K and 5~K, a similar mechanism should be suspected at both temperatures; proposed models must support this scaling while also providing for heating-rate values to vary substantially in amplitude over this temperature range.  The model based on lossy dielectric layers is still viable, and targeted surface preparation and materials characterization in conjunction with careful heating-rate measurements may further elucidate the viability of this or other mechanisms of anomalous ion heating.

\section{Acknowledgments}
We thank Vladimir Bolkhovsky and Sergey Tolpygo for trap fabrication and discussions on film morphology, George Fitch for layout assistance, and Peter Murphy, Chris Thoummaraj, and Karen Magoon for assistance with chip packaging. This work was sponsored by the Assistant Secretary of Defense for Research and Engineering under Air Force contract number FA8721-05-C-0002. Opinions, interpretations, conclusions, and recommendations are those of the authors and are not necessarily endorsed by the United States Government.

\bibliography{distancePaperBib}

\begin{thebibliography}{35}%
\makeatletter
\providecommand \@ifxundefined [1]{%
 \@ifx{#1\undefined}
}%
\providecommand \@ifnum [1]{%
 \ifnum #1\expandafter \@firstoftwo
 \else \expandafter \@secondoftwo
 \fi
}%
\providecommand \@ifx [1]{%
 \ifx #1\expandafter \@firstoftwo
 \else \expandafter \@secondoftwo
 \fi
}%
\providecommand \natexlab [1]{#1}%
\providecommand \enquote  [1]{``#1''}%
\providecommand \bibnamefont  [1]{#1}%
\providecommand \bibfnamefont [1]{#1}%
\providecommand \citenamefont [1]{#1}%
\providecommand \href@noop [0]{\@secondoftwo}%
\providecommand \href [0]{\begingroup \@sanitize@url \@href}%
\providecommand \@href[1]{\@@startlink{#1}\@@href}%
\providecommand \@@href[1]{\endgroup#1\@@endlink}%
\providecommand \@sanitize@url [0]{\catcode `\\12\catcode `\$12\catcode
  `\&12\catcode `\#12\catcode `\^12\catcode `\_12\catcode `\%12\relax}%
\providecommand \@@startlink[1]{}%
\providecommand \@@endlink[0]{}%
\providecommand \url  [0]{\begingroup\@sanitize@url \@url }%
\providecommand \@url [1]{\endgroup\@href {#1}{\urlprefix }}%
\providecommand \urlprefix  [0]{URL }%
\providecommand \Eprint [0]{\href }%
\providecommand \doibase [0]{http://dx.doi.org/}%
\providecommand \selectlanguage [0]{\@gobble}%
\providecommand \bibinfo  [0]{\@secondoftwo}%
\providecommand \bibfield  [0]{\@secondoftwo}%
\providecommand \translation [1]{[#1]}%
\providecommand \BibitemOpen [0]{}%
\providecommand \bibitemStop [0]{}%
\providecommand \bibitemNoStop [0]{.\EOS\space}%
\providecommand \EOS [0]{\spacefactor3000\relax}%
\providecommand \BibitemShut  [1]{\csname bibitem#1\endcsname}%
\let\auto@bib@innerbib\@empty
\bibitem [{\citenamefont {Leibfried}\ \emph {et~al.}(2003)\citenamefont
  {Leibfried}, \citenamefont {Blatt}, \citenamefont {Monroe},\ and\
  \citenamefont {Wineland}}]{Leibfried2003}%
  \BibitemOpen
  \bibfield  {author} {\bibinfo {author} {\bibfnamefont {D.}~\bibnamefont
  {Leibfried}}, \bibinfo {author} {\bibfnamefont {R.}~\bibnamefont {Blatt}},
  \bibinfo {author} {\bibfnamefont {C.}~\bibnamefont {Monroe}}, \ and\ \bibinfo
  {author} {\bibfnamefont {D.}~\bibnamefont {Wineland}},\ }\bibfield  {title}
  {\enquote {\bibinfo {title} {Quantum dynamics of single trapped ions},}\
  }\href {\doibase 10.1103/RevModPhys.75.281} {\bibfield  {journal} {\bibinfo
  {journal} {Rev. Mod. Phys.}\ }\textbf {\bibinfo {volume} {75}},\ \bibinfo
  {pages} {281--324} (\bibinfo {year} {2003})}\BibitemShut {NoStop}%
\bibitem [{\citenamefont {Wineland}\ \emph {et~al.}(1998)\citenamefont
  {Wineland}, \citenamefont {Monroe}, \citenamefont {Itano}, \citenamefont
  {Leibfried}, \citenamefont {King},\ and\ \citenamefont {Meekhof}}]{theBible}%
  \BibitemOpen
  \bibfield  {author} {\bibinfo {author} {\bibfnamefont {D.~J.}\ \bibnamefont
  {Wineland}}, \bibinfo {author} {\bibfnamefont {C.}~\bibnamefont {Monroe}},
  \bibinfo {author} {\bibfnamefont {W.~M.}\ \bibnamefont {Itano}}, \bibinfo
  {author} {\bibfnamefont {D.}~\bibnamefont {Leibfried}}, \bibinfo {author}
  {\bibfnamefont {B.~E.}\ \bibnamefont {King}}, \ and\ \bibinfo {author}
  {\bibfnamefont {D.~M.}\ \bibnamefont {Meekhof}},\ }\bibfield  {title}
  {\enquote {\bibinfo {title} {Experimental issues in coherent quantum-state
  manipulation of trapped atomic ions},}\ }\href {\doibase
  10.6028/jres.103.019} {\bibfield  {journal} {\bibinfo  {journal} {J. Res.
  Natl. Inst. Stand. Technol.}\ }\textbf {\bibinfo {volume} {103}},\ \bibinfo
  {pages} {259--328} (\bibinfo {year} {1998})}\BibitemShut {NoStop}%
\bibitem [{\citenamefont {Turchette}\ \emph {et~al.}(2000)\citenamefont
  {Turchette}, \citenamefont {Kielpinski}, \citenamefont {King}, \citenamefont
  {Leibfried}, \citenamefont {Meekhof}, \citenamefont {Myatt}, \citenamefont
  {Rowe}, \citenamefont {Sackett}, \citenamefont {Wood}, \citenamefont {Itano},
  \citenamefont {Monroe},\ and\ \citenamefont {Wineland}}]{Turchette2000}%
  \BibitemOpen
  \bibfield  {author} {\bibinfo {author} {\bibfnamefont {Q.~A.}\ \bibnamefont
  {Turchette}}, \bibinfo {author} {\bibnamefont {Kielpinski}}, \bibinfo
  {author} {\bibfnamefont {B.~E.}\ \bibnamefont {King}}, \bibinfo {author}
  {\bibfnamefont {D.}~\bibnamefont {Leibfried}}, \bibinfo {author}
  {\bibfnamefont {D.~M.}\ \bibnamefont {Meekhof}}, \bibinfo {author}
  {\bibfnamefont {C.~J.}\ \bibnamefont {Myatt}}, \bibinfo {author}
  {\bibfnamefont {M.~A.}\ \bibnamefont {Rowe}}, \bibinfo {author}
  {\bibfnamefont {C.~A.}\ \bibnamefont {Sackett}}, \bibinfo {author}
  {\bibfnamefont {C.~S.}\ \bibnamefont {Wood}}, \bibinfo {author}
  {\bibfnamefont {W.~M.}\ \bibnamefont {Itano}}, \bibinfo {author}
  {\bibfnamefont {C.}~\bibnamefont {Monroe}}, \ and\ \bibinfo {author}
  {\bibfnamefont {D.~J.}\ \bibnamefont {Wineland}},\ }\bibfield  {title}
  {\enquote {\bibinfo {title} {Heating of trapped ions from the quantum ground
  state},}\ }\href {\doibase 10.1103/PhysRevA.61.063418} {\bibfield  {journal}
  {\bibinfo  {journal} {Phys. Rev. A}\ }\textbf {\bibinfo {volume} {61}},\
  \bibinfo {pages} {063418} (\bibinfo {year} {2000})}\BibitemShut {NoStop}%
\bibitem [{\citenamefont {Deslauriers}\ \emph {et~al.}(2006)\citenamefont
  {Deslauriers}, \citenamefont {Olmschenk}, \citenamefont {Stick},
  \citenamefont {Hensinger}, \citenamefont {Sterk},\ and\ \citenamefont
  {Monroe}}]{PhysRevLett.97.103007}%
  \BibitemOpen
  \bibfield  {author} {\bibinfo {author} {\bibfnamefont {L.}~\bibnamefont
  {Deslauriers}}, \bibinfo {author} {\bibfnamefont {S.}~\bibnamefont
  {Olmschenk}}, \bibinfo {author} {\bibfnamefont {D.}~\bibnamefont {Stick}},
  \bibinfo {author} {\bibfnamefont {W.~K.}\ \bibnamefont {Hensinger}}, \bibinfo
  {author} {\bibfnamefont {J.}~\bibnamefont {Sterk}}, \ and\ \bibinfo {author}
  {\bibfnamefont {C.}~\bibnamefont {Monroe}},\ }\bibfield  {title} {\enquote
  {\bibinfo {title} {Scaling and suppression of anomalous heating in ion
  traps},}\ }\href {\doibase 10.1103/PhysRevLett.97.103007} {\bibfield
  {journal} {\bibinfo  {journal} {Phys. Rev. Lett.}\ }\textbf {\bibinfo
  {volume} {97}},\ \bibinfo {pages} {103007} (\bibinfo {year}
  {2006})}\BibitemShut {NoStop}%
\bibitem [{\citenamefont {Hite}\ \emph {et~al.}(2017)\citenamefont {Hite},
  \citenamefont {McKay}, \citenamefont {Kotler}, \citenamefont {Leibfried},
  \citenamefont {Wineland},\ and\ \citenamefont
  {Pappas}}]{hite_mckay_kotler_leibfried_wineland_pappas_2017}%
  \BibitemOpen
  \bibfield  {author} {\bibinfo {author} {\bibfnamefont {D.~A.}\ \bibnamefont
  {Hite}}, \bibinfo {author} {\bibfnamefont {K.~S.}\ \bibnamefont {McKay}},
  \bibinfo {author} {\bibfnamefont {S.}~\bibnamefont {Kotler}}, \bibinfo
  {author} {\bibfnamefont {D.}~\bibnamefont {Leibfried}}, \bibinfo {author}
  {\bibfnamefont {D.~J.}\ \bibnamefont {Wineland}}, \ and\ \bibinfo {author}
  {\bibfnamefont {D.~P.}\ \bibnamefont {Pappas}},\ }\bibfield  {title}
  {\enquote {\bibinfo {title} {Measurements of trapped-ion heating rates with
  exchangeable surfaces in close proximity},}\ }\href {\doibase
  10.1557/adv.2017.14} {\bibfield  {journal} {\bibinfo  {journal} {MRS
  Advances}\ }\textbf {\bibinfo {volume} {2}},\ \bibinfo {pages} {2189–2197}
  (\bibinfo {year} {2017})}\BibitemShut {NoStop}%
\bibitem [{\citenamefont {Boldin}\ \emph {et~al.}(2017)\citenamefont {Boldin},
  \citenamefont {Kraft},\ and\ \citenamefont
  {Wunderlich}}]{wunderlich_arXiv_2017}%
  \BibitemOpen
  \bibfield  {author} {\bibinfo {author} {\bibfnamefont {I.~A.}\ \bibnamefont
  {Boldin}}, \bibinfo {author} {\bibfnamefont {A.}~\bibnamefont {Kraft}}, \
  and\ \bibinfo {author} {\bibfnamefont {C.}~\bibnamefont {Wunderlich}},\
  }\bibfield  {title} {\enquote {\bibinfo {title} {Measuring anomalous heating
  in a planar ion trap with variable ion-surface separation},}\ }\href@noop {}
  {\  (\bibinfo {year} {2017})},\ \Eprint
  {http://arxiv.org/abs/arXiv:1708.03147} {arXiv:1708.03147} \BibitemShut
  {NoStop}%
\bibitem [{\citenamefont {Benhelm}\ \emph {et~al.}(2008)\citenamefont
  {Benhelm}, \citenamefont {Kirchmair}, \citenamefont {Roos},\ and\
  \citenamefont {Blatt}}]{Inns:HiFi2qubit:NatPhys:08}%
  \BibitemOpen
  \bibfield  {author} {\bibinfo {author} {\bibfnamefont {J.}~\bibnamefont
  {Benhelm}}, \bibinfo {author} {\bibfnamefont {G.}~\bibnamefont {Kirchmair}},
  \bibinfo {author} {\bibfnamefont {C.~F.}\ \bibnamefont {Roos}}, \ and\
  \bibinfo {author} {\bibfnamefont {R.}~\bibnamefont {Blatt}},\ }\bibfield
  {title} {\enquote {\bibinfo {title} {Towards fault-tolerant quantum computing
  with trapped ions},}\ }\href@noop {} {\bibfield  {journal} {\bibinfo
  {journal} {Nature Phys.}\ }\textbf {\bibinfo {volume} {4}},\ \bibinfo {pages}
  {463} (\bibinfo {year} {2008})}\BibitemShut {NoStop}%
\bibitem [{\citenamefont {Ballance}\ \emph {et~al.}(2016)\citenamefont
  {Ballance}, \citenamefont {Harty}, \citenamefont {Linke}, \citenamefont
  {Sepiol},\ and\ \citenamefont {Lucas}}]{PhysRevLett.117.060504}%
  \BibitemOpen
  \bibfield  {author} {\bibinfo {author} {\bibfnamefont {C.~J.}\ \bibnamefont
  {Ballance}}, \bibinfo {author} {\bibfnamefont {T.~P.}\ \bibnamefont {Harty}},
  \bibinfo {author} {\bibfnamefont {N.~M.}\ \bibnamefont {Linke}}, \bibinfo
  {author} {\bibfnamefont {M.~A.}\ \bibnamefont {Sepiol}}, \ and\ \bibinfo
  {author} {\bibfnamefont {D.~M.}\ \bibnamefont {Lucas}},\ }\bibfield  {title}
  {\enquote {\bibinfo {title} {High-fidelity quantum logic gates using
  trapped-ion hyperfine qubits},}\ }\href {\doibase
  10.1103/PhysRevLett.117.060504} {\bibfield  {journal} {\bibinfo  {journal}
  {Phys. Rev. Lett.}\ }\textbf {\bibinfo {volume} {117}},\ \bibinfo {pages}
  {060504} (\bibinfo {year} {2016})}\BibitemShut {NoStop}%
\bibitem [{\citenamefont {Gaebler}\ \emph {et~al.}(2016)\citenamefont
  {Gaebler}, \citenamefont {Tan}, \citenamefont {Lin}, \citenamefont {Wan},
  \citenamefont {Bowler}, \citenamefont {Keith}, \citenamefont {Glancy},
  \citenamefont {Coakley}, \citenamefont {Knill}, \citenamefont {Leibfried},\
  and\ \citenamefont {Wineland}}]{PhysRevLett.117.060505}%
  \BibitemOpen
  \bibfield  {author} {\bibinfo {author} {\bibfnamefont {J.~P.}\ \bibnamefont
  {Gaebler}}, \bibinfo {author} {\bibfnamefont {T.~R.}\ \bibnamefont {Tan}},
  \bibinfo {author} {\bibfnamefont {Y.}~\bibnamefont {Lin}}, \bibinfo {author}
  {\bibfnamefont {Y.}~\bibnamefont {Wan}}, \bibinfo {author} {\bibfnamefont
  {R.}~\bibnamefont {Bowler}}, \bibinfo {author} {\bibfnamefont {A.~C.}\
  \bibnamefont {Keith}}, \bibinfo {author} {\bibfnamefont {S.}~\bibnamefont
  {Glancy}}, \bibinfo {author} {\bibfnamefont {K.}~\bibnamefont {Coakley}},
  \bibinfo {author} {\bibfnamefont {E.}~\bibnamefont {Knill}}, \bibinfo
  {author} {\bibfnamefont {D.}~\bibnamefont {Leibfried}}, \ and\ \bibinfo
  {author} {\bibfnamefont {D.~J.}\ \bibnamefont {Wineland}},\ }\bibfield
  {title} {\enquote {\bibinfo {title} {High-fidelity universal gate set for
  ${^{9}\mathrm{Be}}^{+}$ ion qubits},}\ }\href {\doibase
  10.1103/PhysRevLett.117.060505} {\bibfield  {journal} {\bibinfo  {journal}
  {Phys. Rev. Lett.}\ }\textbf {\bibinfo {volume} {117}},\ \bibinfo {pages}
  {060505} (\bibinfo {year} {2016})}\BibitemShut {NoStop}%
\bibitem [{\citenamefont {Harty}\ \emph {et~al.}(2016)\citenamefont {Harty},
  \citenamefont {Sepiol}, \citenamefont {Allcock}, \citenamefont {Ballance},
  \citenamefont {Tarlton},\ and\ \citenamefont
  {Lucas}}]{PhysRevLett.117.140501}%
  \BibitemOpen
  \bibfield  {author} {\bibinfo {author} {\bibfnamefont {T.~P.}\ \bibnamefont
  {Harty}}, \bibinfo {author} {\bibfnamefont {M.~A.}\ \bibnamefont {Sepiol}},
  \bibinfo {author} {\bibfnamefont {D.~T.~C.}\ \bibnamefont {Allcock}},
  \bibinfo {author} {\bibfnamefont {C.~J.}\ \bibnamefont {Ballance}}, \bibinfo
  {author} {\bibfnamefont {J.~E.}\ \bibnamefont {Tarlton}}, \ and\ \bibinfo
  {author} {\bibfnamefont {D.~M.}\ \bibnamefont {Lucas}},\ }\bibfield  {title}
  {\enquote {\bibinfo {title} {High-fidelity trapped-ion quantum logic using
  near-field microwaves},}\ }\href {\doibase 10.1103/PhysRevLett.117.140501}
  {\bibfield  {journal} {\bibinfo  {journal} {Phys. Rev. Lett.}\ }\textbf
  {\bibinfo {volume} {117}},\ \bibinfo {pages} {140501} (\bibinfo {year}
  {2016})}\BibitemShut {NoStop}%
\bibitem [{\citenamefont {Mehta}(2017)}]{KaranMehtaThesis}%
  \BibitemOpen
  \bibfield  {author} {\bibinfo {author} {\bibfnamefont {K.~K.}\ \bibnamefont
  {Mehta}},\ }\emph {\bibinfo {title} {Integrated optical quantum manipulation
  and measurement of trapped ions}},\ \href@noop {} {Ph.D. thesis},\ \bibinfo
  {school} {Massachusetts Institute of Technology} (\bibinfo {year}
  {2017})\BibitemShut {NoStop}%
\bibitem [{\citenamefont {Slichter}\ \emph {et~al.}(2017)\citenamefont
  {Slichter}, \citenamefont {Verma}, \citenamefont {Leibfried}, \citenamefont
  {Mirin}, \citenamefont {Nam},\ and\ \citenamefont {Wineland}}]{Slichter:17}%
  \BibitemOpen
  \bibfield  {author} {\bibinfo {author} {\bibfnamefont {D.~H.}\ \bibnamefont
  {Slichter}}, \bibinfo {author} {\bibfnamefont {V.~B.}\ \bibnamefont {Verma}},
  \bibinfo {author} {\bibfnamefont {D.}~\bibnamefont {Leibfried}}, \bibinfo
  {author} {\bibfnamefont {R.~P.}\ \bibnamefont {Mirin}}, \bibinfo {author}
  {\bibfnamefont {S.~W.}\ \bibnamefont {Nam}}, \ and\ \bibinfo {author}
  {\bibfnamefont {D.~J.}\ \bibnamefont {Wineland}},\ }\bibfield  {title}
  {\enquote {\bibinfo {title} {Uv-sensitive superconducting nanowire single
  photon detectors for integration in an ion trap},}\ }\href {\doibase
  10.1364/OE.25.008705} {\bibfield  {journal} {\bibinfo  {journal} {Opt.
  Express}\ }\textbf {\bibinfo {volume} {25}},\ \bibinfo {pages} {8705--8720}
  (\bibinfo {year} {2017})}\BibitemShut {NoStop}%
\bibitem [{\citenamefont {Brownnutt}\ \emph {et~al.}(2015)\citenamefont
  {Brownnutt}, \citenamefont {Kumph}, \citenamefont {Rabl},\ and\ \citenamefont
  {Blatt}}]{Brownnutt2015}%
  \BibitemOpen
  \bibfield  {author} {\bibinfo {author} {\bibfnamefont {M.}~\bibnamefont
  {Brownnutt}}, \bibinfo {author} {\bibfnamefont {M.}~\bibnamefont {Kumph}},
  \bibinfo {author} {\bibfnamefont {P.}~\bibnamefont {Rabl}}, \ and\ \bibinfo
  {author} {\bibfnamefont {R.}~\bibnamefont {Blatt}},\ }\bibfield  {title}
  {\enquote {\bibinfo {title} {Ion-trap measurements of electric-field noise
  near surfaces},}\ }\href {\doibase 10.1103/RevModPhys.87.1419} {\bibfield
  {journal} {\bibinfo  {journal} {Rev. Mod. Phys.}\ }\textbf {\bibinfo {volume}
  {87}},\ \bibinfo {pages} {1419--1482} (\bibinfo {year} {2015})}\BibitemShut
  {NoStop}%
\bibitem [{\citenamefont {Dubessy}\ \emph {et~al.}(2009)\citenamefont
  {Dubessy}, \citenamefont {Coudreau},\ and\ \citenamefont
  {Guidoni}}]{PhysRevA.80.031402}%
  \BibitemOpen
  \bibfield  {author} {\bibinfo {author} {\bibfnamefont {R.}~\bibnamefont
  {Dubessy}}, \bibinfo {author} {\bibfnamefont {T.}~\bibnamefont {Coudreau}}, \
  and\ \bibinfo {author} {\bibfnamefont {L.}~\bibnamefont {Guidoni}},\
  }\bibfield  {title} {\enquote {\bibinfo {title} {Electric field noise above
  surfaces: A model for heating-rate scaling law in ion traps},}\ }\href
  {\doibase 10.1103/PhysRevA.80.031402} {\bibfield  {journal} {\bibinfo
  {journal} {Phys. Rev. A}\ }\textbf {\bibinfo {volume} {80}},\ \bibinfo
  {pages} {031402} (\bibinfo {year} {2009})}\BibitemShut {NoStop}%
\bibitem [{\citenamefont {Low}\ \emph {et~al.}(2011)\citenamefont {Low},
  \citenamefont {Herskind},\ and\ \citenamefont {Chuang}}]{PhysRevA.84.053425}%
  \BibitemOpen
  \bibfield  {author} {\bibinfo {author} {\bibfnamefont {G.~H.}\ \bibnamefont
  {Low}}, \bibinfo {author} {\bibfnamefont {P.~F.}\ \bibnamefont {Herskind}}, \
  and\ \bibinfo {author} {\bibfnamefont {I.~L.}\ \bibnamefont {Chuang}},\
  }\bibfield  {title} {\enquote {\bibinfo {title} {Finite-geometry models of
  electric field noise from patch potentials in ion traps},}\ }\href {\doibase
  10.1103/PhysRevA.84.053425} {\bibfield  {journal} {\bibinfo  {journal} {Phys.
  Rev. A}\ }\textbf {\bibinfo {volume} {84}},\ \bibinfo {pages} {053425}
  (\bibinfo {year} {2011})}\BibitemShut {NoStop}%
\bibitem [{\citenamefont {Chiaverini}\ \emph {et~al.}(2005)\citenamefont
  {Chiaverini}, \citenamefont {Blakestad}, \citenamefont {Britton},
  \citenamefont {Jost}, \citenamefont {Langer}, \citenamefont {Leibfried},
  \citenamefont {Ozeri},\ and\ \citenamefont {Wineland}}]{NIST:SET:QIC:05}%
  \BibitemOpen
  \bibfield  {author} {\bibinfo {author} {\bibfnamefont {J.}~\bibnamefont
  {Chiaverini}}, \bibinfo {author} {\bibfnamefont {R.~B.}\ \bibnamefont
  {Blakestad}}, \bibinfo {author} {\bibfnamefont {J.}~\bibnamefont {Britton}},
  \bibinfo {author} {\bibfnamefont {J.~D.}\ \bibnamefont {Jost}}, \bibinfo
  {author} {\bibfnamefont {C.}~\bibnamefont {Langer}}, \bibinfo {author}
  {\bibfnamefont {D.}~\bibnamefont {Leibfried}}, \bibinfo {author}
  {\bibfnamefont {R.}~\bibnamefont {Ozeri}}, \ and\ \bibinfo {author}
  {\bibfnamefont {D.~J.}\ \bibnamefont {Wineland}},\ }\bibfield  {title}
  {\enquote {\bibinfo {title} {Surface-electrode architecture for ion-trap
  quantum information processing},}\ }\href@noop {} {\bibfield  {journal}
  {\bibinfo  {journal} {Quantum Inf. Comput.}\ }\textbf {\bibinfo {volume}
  {5}},\ \bibinfo {pages} {419} (\bibinfo {year} {2005})}\BibitemShut {NoStop}%
\bibitem [{\citenamefont {Mehta}\ \emph {et~al.}(2016)\citenamefont {Mehta},
  \citenamefont {Bruzewicz}, \citenamefont {McConnell}, \citenamefont {Ram},
  \citenamefont {Sage},\ and\ \citenamefont {Chiaverini}}]{Mehta_NNano_2016}%
  \BibitemOpen
  \bibfield  {author} {\bibinfo {author} {\bibfnamefont {K.~K.}\ \bibnamefont
  {Mehta}}, \bibinfo {author} {\bibfnamefont {C.~D.}\ \bibnamefont
  {Bruzewicz}}, \bibinfo {author} {\bibfnamefont {R.}~\bibnamefont
  {McConnell}}, \bibinfo {author} {\bibfnamefont {R.~J.}\ \bibnamefont {Ram}},
  \bibinfo {author} {\bibfnamefont {J.~M.}\ \bibnamefont {Sage}}, \ and\
  \bibinfo {author} {\bibfnamefont {J.}~\bibnamefont {Chiaverini}},\ }\bibfield
   {title} {\enquote {\bibinfo {title} {Integrated optical addressing of an ion
  qubit},}\ }\href {\doibase 10.1038/nnano.2016.139} {\bibfield  {journal}
  {\bibinfo  {journal} {Nature Nanotechnology}\ }\textbf {\bibinfo {volume}
  {11}},\ \bibinfo {pages} {1066--1070} (\bibinfo {year} {2016})}\BibitemShut
  {NoStop}%
\bibitem [{\citenamefont {Lekitsch}\ \emph {et~al.}(2017)\citenamefont
  {Lekitsch}, \citenamefont {Weidt}, \citenamefont {Fowler}, \citenamefont
  {M{\o}lmer}, \citenamefont {Devitt}, \citenamefont {Wunderlich},\ and\
  \citenamefont {Hensinger}}]{Lekitsch2017}%
  \BibitemOpen
  \bibfield  {author} {\bibinfo {author} {\bibfnamefont {B.}~\bibnamefont
  {Lekitsch}}, \bibinfo {author} {\bibfnamefont {S.}~\bibnamefont {Weidt}},
  \bibinfo {author} {\bibfnamefont {A.~G.}\ \bibnamefont {Fowler}}, \bibinfo
  {author} {\bibfnamefont {K.}~\bibnamefont {M{\o}lmer}}, \bibinfo {author}
  {\bibfnamefont {S.~J.}\ \bibnamefont {Devitt}}, \bibinfo {author}
  {\bibfnamefont {C.}~\bibnamefont {Wunderlich}}, \ and\ \bibinfo {author}
  {\bibfnamefont {W.~K.}\ \bibnamefont {Hensinger}},\ }\bibfield  {title}
  {\enquote {\bibinfo {title} {Blueprint for a microwave trapped ion quantum
  computer},}\ }\href {\doibase 10.1126/sciadv.1601540} {\bibfield  {journal}
  {\bibinfo  {journal} {Sci. Adv.}\ }\textbf {\bibinfo {volume} {3}},\ \bibinfo
  {pages} {e1601540} (\bibinfo {year} {2017})}\BibitemShut {NoStop}%
\bibitem [{\citenamefont {Monroe}\ \emph {et~al.}(1995)\citenamefont {Monroe},
  \citenamefont {Meekhof}, \citenamefont {King}, \citenamefont {Jefferts},
  \citenamefont {Itano}, \citenamefont {Wineland},\ and\ \citenamefont
  {Gould}}]{PhysRevLett.75.4011}%
  \BibitemOpen
  \bibfield  {author} {\bibinfo {author} {\bibfnamefont {C.}~\bibnamefont
  {Monroe}}, \bibinfo {author} {\bibfnamefont {D.~M.}\ \bibnamefont {Meekhof}},
  \bibinfo {author} {\bibfnamefont {B.~E.}\ \bibnamefont {King}}, \bibinfo
  {author} {\bibfnamefont {S.~R.}\ \bibnamefont {Jefferts}}, \bibinfo {author}
  {\bibfnamefont {W.~M.}\ \bibnamefont {Itano}}, \bibinfo {author}
  {\bibfnamefont {D.~J.}\ \bibnamefont {Wineland}}, \ and\ \bibinfo {author}
  {\bibfnamefont {P.}~\bibnamefont {Gould}},\ }\bibfield  {title} {\enquote
  {\bibinfo {title} {Resolved-sideband raman cooling of a bound atom to the 3d
  zero-point energy},}\ }\href {\doibase 10.1103/PhysRevLett.75.4011}
  {\bibfield  {journal} {\bibinfo  {journal} {Phys. Rev. Lett.}\ }\textbf
  {\bibinfo {volume} {75}},\ \bibinfo {pages} {4011--4014} (\bibinfo {year}
  {1995})}\BibitemShut {NoStop}%
\bibitem [{\citenamefont {Rowe}\ \emph {et~al.}(2002)\citenamefont {Rowe},
  \citenamefont {Ben-Kish}, \citenamefont {DeMarco}, \citenamefont {Leibfried},
  \citenamefont {Meyer}, \citenamefont {Beall}, \citenamefont {Britton},
  \citenamefont {Hughes}, \citenamefont {Itano}, \citenamefont {Jelenkovic},
  \citenamefont {Langer}, \citenamefont {Rosenband},\ and\ \citenamefont
  {Wineland}}]{RoweQIC2002}%
  \BibitemOpen
  \bibfield  {author} {\bibinfo {author} {\bibfnamefont {M.~A.}\ \bibnamefont
  {Rowe}}, \bibinfo {author} {\bibfnamefont {A.}~\bibnamefont {Ben-Kish}},
  \bibinfo {author} {\bibfnamefont {B.}~\bibnamefont {DeMarco}}, \bibinfo
  {author} {\bibfnamefont {D.}~\bibnamefont {Leibfried}}, \bibinfo {author}
  {\bibfnamefont {V.}~\bibnamefont {Meyer}}, \bibinfo {author} {\bibfnamefont
  {J.}~\bibnamefont {Beall}}, \bibinfo {author} {\bibfnamefont
  {J.}~\bibnamefont {Britton}}, \bibinfo {author} {\bibfnamefont
  {J.}~\bibnamefont {Hughes}}, \bibinfo {author} {\bibfnamefont {W.~M.}\
  \bibnamefont {Itano}}, \bibinfo {author} {\bibfnamefont {B.}~\bibnamefont
  {Jelenkovic}}, \bibinfo {author} {\bibfnamefont {C.}~\bibnamefont {Langer}},
  \bibinfo {author} {\bibfnamefont {T.}~\bibnamefont {Rosenband}}, \ and\
  \bibinfo {author} {\bibfnamefont {D.~J.}\ \bibnamefont {Wineland}},\
  }\bibfield  {title} {\enquote {\bibinfo {title} {Transport of quantum states
  and separation of ions in a dual rf ion trap},}\ }\href@noop {} {\bibfield
  {journal} {\bibinfo  {journal} {Quantum Inf. Comput.}\ }\textbf {\bibinfo
  {volume} {2}},\ \bibinfo {pages} {257--271} (\bibinfo {year}
  {2002})}\BibitemShut {NoStop}%
\bibitem [{\citenamefont {House}(2008)}]{PhysRevA.78.033402}%
  \BibitemOpen
  \bibfield  {author} {\bibinfo {author} {\bibfnamefont {M.~G.}\ \bibnamefont
  {House}},\ }\bibfield  {title} {\enquote {\bibinfo {title} {Analytic model
  for electrostatic fields in surface-electrode ion traps},}\ }\href {\doibase
  10.1103/PhysRevA.78.033402} {\bibfield  {journal} {\bibinfo  {journal} {Phys.
  Rev. A}\ }\textbf {\bibinfo {volume} {78}},\ \bibinfo {pages} {033402}
  (\bibinfo {year} {2008})}\BibitemShut {NoStop}%
\bibitem [{\citenamefont {Wesenberg}(2008)}]{PhysRevA.78.063410}%
  \BibitemOpen
  \bibfield  {author} {\bibinfo {author} {\bibfnamefont {J.~H.}\ \bibnamefont
  {Wesenberg}},\ }\bibfield  {title} {\enquote {\bibinfo {title}
  {Electrostatics of surface-electrode ion traps},}\ }\href {\doibase
  10.1103/PhysRevA.78.063410} {\bibfield  {journal} {\bibinfo  {journal} {Phys.
  Rev. A}\ }\textbf {\bibinfo {volume} {78}},\ \bibinfo {pages} {063410}
  (\bibinfo {year} {2008})}\BibitemShut {NoStop}%
\bibitem [{\citenamefont {Chiaverini}\ and\ \citenamefont
  {Sage}(2014)}]{Chiaverini2014}%
  \BibitemOpen
  \bibfield  {author} {\bibinfo {author} {\bibfnamefont {J.}~\bibnamefont
  {Chiaverini}}\ and\ \bibinfo {author} {\bibfnamefont {J.~M.}\ \bibnamefont
  {Sage}},\ }\bibfield  {title} {\enquote {\bibinfo {title} {Insensitivity of
  the rate of ion motional heating to trap-electrode material over a large
  temperature range},}\ }\href {\doibase 10.1103/PhysRevA.89.012318} {\bibfield
   {journal} {\bibinfo  {journal} {Phys. Rev. A}\ }\textbf {\bibinfo {volume}
  {89}},\ \bibinfo {pages} {012318} (\bibinfo {year} {2014})}\BibitemShut
  {NoStop}%
\bibitem [{\citenamefont {Bruzewicz}\ \emph {et~al.}(2016)\citenamefont
  {Bruzewicz}, \citenamefont {McConnell}, \citenamefont {Chiaverini},\ and\
  \citenamefont {Sage}}]{bruzewicz2016scalable}%
  \BibitemOpen
  \bibfield  {author} {\bibinfo {author} {\bibfnamefont {C.~D.}\ \bibnamefont
  {Bruzewicz}}, \bibinfo {author} {\bibfnamefont {R.}~\bibnamefont
  {McConnell}}, \bibinfo {author} {\bibfnamefont {J.}~\bibnamefont
  {Chiaverini}}, \ and\ \bibinfo {author} {\bibfnamefont {J.~M}\ \bibnamefont
  {Sage}},\ }\bibfield  {title} {\enquote {\bibinfo {title} {Scalable loading
  of a two-dimensional trapped-ion array},}\ }\href {\doibase
  10.1038/ncomms13005} {\bibfield  {journal} {\bibinfo  {journal} {Nat.
  Commun.}\ }\textbf {\bibinfo {volume} {7}},\ \bibinfo {pages} {13005}
  (\bibinfo {year} {2016})}\BibitemShut {NoStop}%
\bibitem [{\citenamefont {Shu}\ \emph {et~al.}(2014)\citenamefont {Shu},
  \citenamefont {Vittorini}, \citenamefont {Buikema}, \citenamefont {Nichols},
  \citenamefont {Volin}, \citenamefont {Stick},\ and\ \citenamefont
  {Brown}}]{PhysRevA.89.062308}%
  \BibitemOpen
  \bibfield  {author} {\bibinfo {author} {\bibfnamefont {G.}~\bibnamefont
  {Shu}}, \bibinfo {author} {\bibfnamefont {G.}~\bibnamefont {Vittorini}},
  \bibinfo {author} {\bibfnamefont {A.}~\bibnamefont {Buikema}}, \bibinfo
  {author} {\bibfnamefont {C.~S.}\ \bibnamefont {Nichols}}, \bibinfo {author}
  {\bibfnamefont {C.}~\bibnamefont {Volin}}, \bibinfo {author} {\bibfnamefont
  {D.}~\bibnamefont {Stick}}, \ and\ \bibinfo {author} {\bibfnamefont
  {Kenneth~R.}\ \bibnamefont {Brown}},\ }\bibfield  {title} {\enquote {\bibinfo
  {title} {Heating rates and ion-motion control in a $\mathsf{Y}$-junction
  surface-electrode trap},}\ }\href {\doibase 10.1103/PhysRevA.89.062308}
  {\bibfield  {journal} {\bibinfo  {journal} {Phys. Rev. A}\ }\textbf {\bibinfo
  {volume} {89}},\ \bibinfo {pages} {062308} (\bibinfo {year}
  {2014})}\BibitemShut {NoStop}%
\bibitem [{\citenamefont {Sedlacek}\ \emph {et~al.}(2017)\citenamefont
  {Sedlacek}, \citenamefont {Stuart}, \citenamefont {Loh}, \citenamefont
  {McConnell}, \citenamefont {Bruzewicz}, \citenamefont {Sage},\ and\
  \citenamefont {Chiaverini}}]{arxiv_noise_paper}%
  \BibitemOpen
  \bibfield  {author} {\bibinfo {author} {\bibfnamefont {J.~A.}\ \bibnamefont
  {Sedlacek}}, \bibinfo {author} {\bibfnamefont {J.}~\bibnamefont {Stuart}},
  \bibinfo {author} {\bibfnamefont {W.}~\bibnamefont {Loh}}, \bibinfo {author}
  {\bibfnamefont {R.}~\bibnamefont {McConnell}}, \bibinfo {author}
  {\bibfnamefont {C.~D.}\ \bibnamefont {Bruzewicz}}, \bibinfo {author}
  {\bibfnamefont {J.~M.}\ \bibnamefont {Sage}}, \ and\ \bibinfo {author}
  {\bibfnamefont {J.}~\bibnamefont {Chiaverini}},\ }\bibfield  {title}
  {\enquote {\bibinfo {title} {Method for determination of technical noise
  limitations to ion motional heating},}\ }\href@noop {} {\  (\bibinfo {year}
  {2017})},\ \Eprint {http://arxiv.org/abs/in preparation} {in preparation}
  \BibitemShut {NoStop}%
\bibitem [{\citenamefont {Blakestad}\ \emph {et~al.}(2009)\citenamefont
  {Blakestad}, \citenamefont {Ospelkaus}, \citenamefont {VanDevender},
  \citenamefont {Amini}, \citenamefont {Britton}, \citenamefont {Leibfried},\
  and\ \citenamefont {Wineland}}]{Blakestad2009}%
  \BibitemOpen
  \bibfield  {author} {\bibinfo {author} {\bibfnamefont {R.~B.}\ \bibnamefont
  {Blakestad}}, \bibinfo {author} {\bibfnamefont {C.}~\bibnamefont
  {Ospelkaus}}, \bibinfo {author} {\bibfnamefont {A.~P.}\ \bibnamefont
  {VanDevender}}, \bibinfo {author} {\bibfnamefont {J.~M.}\ \bibnamefont
  {Amini}}, \bibinfo {author} {\bibfnamefont {J.}~\bibnamefont {Britton}},
  \bibinfo {author} {\bibfnamefont {D.}~\bibnamefont {Leibfried}}, \ and\
  \bibinfo {author} {\bibfnamefont {D.~J.}\ \bibnamefont {Wineland}},\
  }\bibfield  {title} {\enquote {\bibinfo {title} {High-fidelity transport of
  trapped-ion qubits through an $\mathbf{X}$-junction trap array},}\ }\href
  {\doibase 10.1103/PhysRevLett.102.153002} {\bibfield  {journal} {\bibinfo
  {journal} {Phys. Rev. Lett.}\ }\textbf {\bibinfo {volume} {102}},\ \bibinfo
  {pages} {153002} (\bibinfo {year} {2009})}\BibitemShut {NoStop}%
\bibitem [{\citenamefont {Wang}\ \emph {et~al.}(2010)\citenamefont {Wang},
  \citenamefont {Ge}, \citenamefont {Labaziewicz}, \citenamefont {Dauler},
  \citenamefont {Berggren},\ and\ \citenamefont {Chuang}}]{Wang2010}%
  \BibitemOpen
  \bibfield  {author} {\bibinfo {author} {\bibfnamefont {S.~X.}\ \bibnamefont
  {Wang}}, \bibinfo {author} {\bibfnamefont {Y.}~\bibnamefont {Ge}}, \bibinfo
  {author} {\bibfnamefont {J.}~\bibnamefont {Labaziewicz}}, \bibinfo {author}
  {\bibfnamefont {E.}~\bibnamefont {Dauler}}, \bibinfo {author} {\bibfnamefont
  {K.}~\bibnamefont {Berggren}}, \ and\ \bibinfo {author} {\bibfnamefont
  {I.~L.}\ \bibnamefont {Chuang}},\ }\bibfield  {title} {\enquote {\bibinfo
  {title} {Superconducting microfabricated ion traps},}\ }\href {\doibase
  10.1063/1.3526733} {\bibfield  {journal} {\bibinfo  {journal} {Applied
  Physics Letters}\ }\textbf {\bibinfo {volume} {97}},\ \bibinfo {pages}
  {244102} (\bibinfo {year} {2010})}\BibitemShut {NoStop}%
\bibitem [{\citenamefont {Leibrandt}\ \emph {et~al.}(2007)\citenamefont
  {Leibrandt}, \citenamefont {Yurke},\ and\ \citenamefont
  {Slusher}}]{Leibrandt2007}%
  \BibitemOpen
  \bibfield  {author} {\bibinfo {author} {\bibfnamefont {D.}~\bibnamefont
  {Leibrandt}}, \bibinfo {author} {\bibfnamefont {B.}~\bibnamefont {Yurke}}, \
  and\ \bibinfo {author} {\bibfnamefont {R.}~\bibnamefont {Slusher}},\
  }\bibfield  {title} {\enquote {\bibinfo {title} {Modeling ion trap thermal
  noise decoherence},}\ }\href
  {http://dl.acm.org/citation.cfm?id=2011706.2011708} {\bibfield  {journal}
  {\bibinfo  {journal} {Quantum Inf. Comput.}\ }\textbf {\bibinfo {volume}
  {7}},\ \bibinfo {pages} {52--72} (\bibinfo {year} {2007})}\BibitemShut
  {NoStop}%
\bibitem [{\citenamefont {Kim}\ \emph {et~al.}(2017)\citenamefont {Kim},
  \citenamefont {Safavi-Naini}, \citenamefont {Hite}, \citenamefont {McKay},
  \citenamefont {Pappas}, \citenamefont {Weck},\ and\ \citenamefont
  {Sadeghpour}}]{PhysRevA.95.033407}%
  \BibitemOpen
  \bibfield  {author} {\bibinfo {author} {\bibfnamefont {E.}~\bibnamefont
  {Kim}}, \bibinfo {author} {\bibfnamefont {A.}~\bibnamefont {Safavi-Naini}},
  \bibinfo {author} {\bibfnamefont {D.~A.}\ \bibnamefont {Hite}}, \bibinfo
  {author} {\bibfnamefont {K.~S.}\ \bibnamefont {McKay}}, \bibinfo {author}
  {\bibfnamefont {D.~P.}\ \bibnamefont {Pappas}}, \bibinfo {author}
  {\bibfnamefont {P.~F.}\ \bibnamefont {Weck}}, \ and\ \bibinfo {author}
  {\bibfnamefont {H.~R.}\ \bibnamefont {Sadeghpour}},\ }\bibfield  {title}
  {\enquote {\bibinfo {title} {Electric-field noise from carbon-adatom
  diffusion on a au(110) surface: First-principles calculations and
  experiments},}\ }\href {\doibase 10.1103/PhysRevA.95.033407} {\bibfield
  {journal} {\bibinfo  {journal} {Phys. Rev. A}\ }\textbf {\bibinfo {volume}
  {95}},\ \bibinfo {pages} {033407} (\bibinfo {year} {2017})}\BibitemShut
  {NoStop}%
\bibitem [{\citenamefont {Safavi-Naini}\ \emph {et~al.}(2013)\citenamefont
  {Safavi-Naini}, \citenamefont {Kim}, \citenamefont {Weck}, \citenamefont
  {Rabl},\ and\ \citenamefont {Sadeghpour}}]{PhysRevA.87.023421}%
  \BibitemOpen
  \bibfield  {author} {\bibinfo {author} {\bibfnamefont {A.}~\bibnamefont
  {Safavi-Naini}}, \bibinfo {author} {\bibfnamefont {E.}~\bibnamefont {Kim}},
  \bibinfo {author} {\bibfnamefont {P.~F.}\ \bibnamefont {Weck}}, \bibinfo
  {author} {\bibfnamefont {P.}~\bibnamefont {Rabl}}, \ and\ \bibinfo {author}
  {\bibfnamefont {H.~R.}\ \bibnamefont {Sadeghpour}},\ }\bibfield  {title}
  {\enquote {\bibinfo {title} {Influence of monolayer contamination on
  electric-field-noise heating in ion traps},}\ }\href {\doibase
  10.1103/PhysRevA.87.023421} {\bibfield  {journal} {\bibinfo  {journal} {Phys.
  Rev. A}\ }\textbf {\bibinfo {volume} {87}},\ \bibinfo {pages} {023421}
  (\bibinfo {year} {2013})}\BibitemShut {NoStop}%
\bibitem [{\citenamefont {Dutta}\ and\ \citenamefont
  {Horn}(1981)}]{RevModPhys.53.497}%
  \BibitemOpen
  \bibfield  {author} {\bibinfo {author} {\bibfnamefont {P.}~\bibnamefont
  {Dutta}}\ and\ \bibinfo {author} {\bibfnamefont {P.~M.}\ \bibnamefont
  {Horn}},\ }\bibfield  {title} {\enquote {\bibinfo {title} {Low-frequency
  fluctuations in solids: $\frac{1}{f}$ noise},}\ }\href {\doibase
  10.1103/RevModPhys.53.497} {\bibfield  {journal} {\bibinfo  {journal} {Rev.
  Mod. Phys.}\ }\textbf {\bibinfo {volume} {53}},\ \bibinfo {pages} {497--516}
  (\bibinfo {year} {1981})}\BibitemShut {NoStop}%
\bibitem [{\citenamefont {Labaziewicz}\ \emph {et~al.}(2008)\citenamefont
  {Labaziewicz}, \citenamefont {Ge}, \citenamefont {Leibrandt}, \citenamefont
  {Wang}, \citenamefont {Shewmon},\ and\ \citenamefont
  {Chuang}}]{Labaziewicz2008}%
  \BibitemOpen
  \bibfield  {author} {\bibinfo {author} {\bibfnamefont {J.}~\bibnamefont
  {Labaziewicz}}, \bibinfo {author} {\bibfnamefont {Y.}~\bibnamefont {Ge}},
  \bibinfo {author} {\bibfnamefont {D.~R.}\ \bibnamefont {Leibrandt}}, \bibinfo
  {author} {\bibfnamefont {S.~X.}\ \bibnamefont {Wang}}, \bibinfo {author}
  {\bibfnamefont {R.}~\bibnamefont {Shewmon}}, \ and\ \bibinfo {author}
  {\bibfnamefont {I.~L.}\ \bibnamefont {Chuang}},\ }\bibfield  {title}
  {\enquote {\bibinfo {title} {Temperature dependence of electric field noise
  above gold surfaces},}\ }\href {\doibase 10.1103/PhysRevLett.101.180602}
  {\bibfield  {journal} {\bibinfo  {journal} {Phys. Rev. Lett.}\ }\textbf
  {\bibinfo {volume} {101}},\ \bibinfo {pages} {180602} (\bibinfo {year}
  {2008})}\BibitemShut {NoStop}%
\bibitem [{\citenamefont {Bruzewicz}\ \emph {et~al.}(2015)\citenamefont
  {Bruzewicz}, \citenamefont {Sage},\ and\ \citenamefont
  {Chiaverini}}]{Bruzewicz2015}%
  \BibitemOpen
  \bibfield  {author} {\bibinfo {author} {\bibfnamefont {C.~D.}\ \bibnamefont
  {Bruzewicz}}, \bibinfo {author} {\bibfnamefont {J.~M.}\ \bibnamefont {Sage}},
  \ and\ \bibinfo {author} {\bibfnamefont {J.}~\bibnamefont {Chiaverini}},\
  }\bibfield  {title} {\enquote {\bibinfo {title} {Measurement of ion motional
  heating rates over a range of trap frequencies and temperatures},}\ }\href
  {\doibase 10.1103/PhysRevA.91.041402} {\bibfield  {journal} {\bibinfo
  {journal} {Phys. Rev. A}\ }\textbf {\bibinfo {volume} {91}},\ \bibinfo
  {pages} {041402} (\bibinfo {year} {2015})}\BibitemShut {NoStop}%
\bibitem [{\citenamefont {Kumph}\ \emph {et~al.}(2016)\citenamefont {Kumph},
  \citenamefont {Henkel}, \citenamefont {Rabl}, \citenamefont {Brownnutt},\
  and\ \citenamefont {Blatt}}]{kumph_NJP_2016}%
  \BibitemOpen
  \bibfield  {author} {\bibinfo {author} {\bibfnamefont {M.}~\bibnamefont
  {Kumph}}, \bibinfo {author} {\bibfnamefont {C.}~\bibnamefont {Henkel}},
  \bibinfo {author} {\bibfnamefont {P.}~\bibnamefont {Rabl}}, \bibinfo {author}
  {\bibfnamefont {M.}~\bibnamefont {Brownnutt}}, \ and\ \bibinfo {author}
  {\bibfnamefont {R.}~\bibnamefont {Blatt}},\ }\bibfield  {title} {\enquote
  {\bibinfo {title} {Electric-field noise above a thin dielectric layer on
  metal electrodes},}\ }\href {http://stacks.iop.org/1367-2630/18/i=2/a=023020}
  {\bibfield  {journal} {\bibinfo  {journal} {New Journal of Physics}\ }\textbf
  {\bibinfo {volume} {18}},\ \bibinfo {pages} {023020} (\bibinfo {year}
  {2016})}\BibitemShut {NoStop}%
\end{thebibliography}%

\end{document}